\documentclass[10pt,final,journal]{IEEEtran}
\usepackage{latexsym}
\usepackage{epsfig}
\usepackage{color}
\usepackage{amsmath} % real numbers, natural numbers, ...
\usepackage{amssymb}
\usepackage{amsxtra}
\usepackage{enumitem}

\usepackage[T1]{fontenc}% optional T1 font encoding
\usepackage{setspace}

\usepackage{amsmath}
\usepackage[cmintegrals]{newtxmath}
\usepackage{graphicx}
\graphicspath{{Figures/}}
\usepackage{url}
\usepackage{cite}
\usepackage{algorithm}
\usepackage{algpseudocode}
\usepackage{subcaption}

\makeatletter
\def\BState{\State\hskip-\ALG@thistlm}
\makeatother

\usepackage{epstopdf}
\usepackage{amsmath}
\usepackage{amsxtra}
\usepackage{amstext}
\usepackage{amssymb}
\usepackage{latexsym}
\usepackage{dsfont} % for \mathds{N}
\usepackage{color}
\usepackage{multirow}
\usepackage{float}
\usepackage{microtype}

\usepackage{tabularx,colortbl}
\usepackage[table]{xcolor}
\usepackage{lscape}
\usepackage{algorithm}
\usepackage{algpseudocode}

\usepackage{cuted}
\usepackage{units}
\usepackage{cases}

\graphicspath{{Figures/}}

\usepackage{mathtools}
\mathtoolsset{showonlyrefs} %only reference numbered equations

\usepackage{tikz}
\usetikzlibrary{shapes}
\usepackage{filecontents}
\usepackage{csvsimple}
\usepackage{pgfplots} % loads tikz
\newcommand{\mbf}[1]{\mathbf{#1}}
\newcommand{\green}[1]{{\color{green}{#1}}}

\usepackage{mathtools}

\begin{document}

\title{Bridging the complexity gap in Tbps-achieving THz-band baseband processing %via channel-induced baseband parallelizability
%Bridging the Tbps gap of THz communications via channel-induced baseband parallelizability
%Bridging the Tbps gap of THz communications: Parallelizability and pseudo-soft information
%Leveraging parallelizability and pseudo-soft information for bridging the THz-band Tbps gap
}

\author{Hadi~Sarieddeen,~\IEEEmembership{Member,~IEEE,}
    Hakim~Jemaa,~\IEEEmembership{Student Member,~IEEE,}
    Simon~Tarboush,
    Christoph~Studer,~\IEEEmembership{Senior Member,~IEEE,}
    Mohamed-Slim~Alouini,~\IEEEmembership{Fellow,~IEEE,}
    and Tareq~Y.~Al-Naffouri,~\IEEEmembership{Senior Member,~IEEE}
\thanks{
H. Sarieddeen is with the Electrical and Computer Engineering Department, American University of Beirut (AUB), Lebanon (hadi.sarieddeen@aub.edu.lb).
H. Jemaa, M.-S. Alouini and T. Al-Naffouri are with the Department of Computer, Electrical and Mathematical Sciences and Engineering (CEMSE), King Abdullah University of Science and Technology (KAUST), Thuwal, Makkah Province, Kingdom of Saudi Arabia, 23955-6900 (hakim.jemaa@kaust.edu.sa; slim.alouini@kaust.edu.sa; tareq.alnaffouri@kaust.edu.sa). S. Tarboush is a researcher from Damascus, Syria (simon.w.tarboush@gmail.com). C. Studer is with the Department of Information Technology and Electrical Engineering, ETH Zurich, Switzerland (studer@ethz.ch).

The first three authors are co-first authors; they contributed equally to this paper. This work was supported by the KAUST Office of Sponsored Research under Award ORA-CRG2021-4695, and the AUB University Research Board.}% <-this % stops a space
}

\maketitle

\begin{abstract}

Recent advances in electronic and photonic technologies have allowed efficient signal generation and transmission at terahertz (THz) frequencies. However, as the gap in THz-operating devices narrows, the demand for terabit-per-second (Tbps)-achieving circuits is increasing. Translating the available hundreds of gigahertz (GHz) of bandwidth into a Tbps data rate requires processing thousands of information bits per clock cycle at state-of-the-art clock frequencies of digital baseband processing circuitry of a few GHz. This paper addresses these constraints and emphasizes the importance of parallelization in signal processing, particularly for channel code decoding. By leveraging structured sub-spaces of THz channels, we propose mapping bits to transmission resources using shorter code-words, extending parallelizability across all baseband processing blocks. THz channels exhibit quasi-deterministic frequency, time, and space structures that enable efficient parallel bit mapping at the source and provide pseudo-soft bit reliability information for efficient detection and decoding at the receiver.
%, eliminating the need for interleavers and complex soft detection and decoding schemes.

\end{abstract}

\section{Introduction}
\label{sec:introduction}

\IEEEPARstart{T}{erahertz} (THz)-band communications over carrier frequencies between $\unit[0.1]{THz}$ and $\unit[10]{THz}$ hold great potential for diverse applications in future generations of wireless communications. The abundantly available contiguous bandwidths at the THz band can support terabit-per-second (Tbps) peak data rates, surpassing the capabilities of millimeter-wave (mmWave) and conventional sub-\unit[6]{GHz} communication systems. However, realizing the full potential of THz bandwidths requires overcoming technological challenges at the device level, often referred to as the ``THz gap''~\cite{akyildiz2022terahertz}. Another significant limitation of THz communications is the reduced communication distance due to spreading and molecular absorption losses. Infrastructure elements like reconfigurable intelligent surfaces (RISs) and ultra-massive multiple-input multiple-output (UM-MIMO) antenna arrays can be utilized to create synthetic propagation paths and achieve beamforming gains, extending the communication range and improving overall performance~\cite{sarieddeen2021overview}.

% chen2022tutorial

% The challenges in the physical layer involve waveform and modulation design~\cite{tarboush2022single} and transceiver architectures~\cite{han2021hybrid}. 

As more THz communication bandwidths get conquered, more stringent baseband processing constraints emerge~\cite{sarieddeen2021overview}. Achieving a Tbps data rate necessitates parallelizable transceiver operations that meet hardware limitations in data conversion sampling frequencies and digital integrated circuit clock frequencies. This paper proposes a framework for efficient THz-specific signal processing to bridge the emerging ``Tbps gap'' that we introduce in Sec.~\ref{sec:tbps_THz}. The unique characteristics of the THz channel drive the exploration of novel low-complexity baseband signal processing techniques, where channel code decoding and UM-MIMO data detection are particularly computationally demanding. 

\begin{figure*}[t]
  \centering
  \includegraphics[width = 0.97\linewidth]{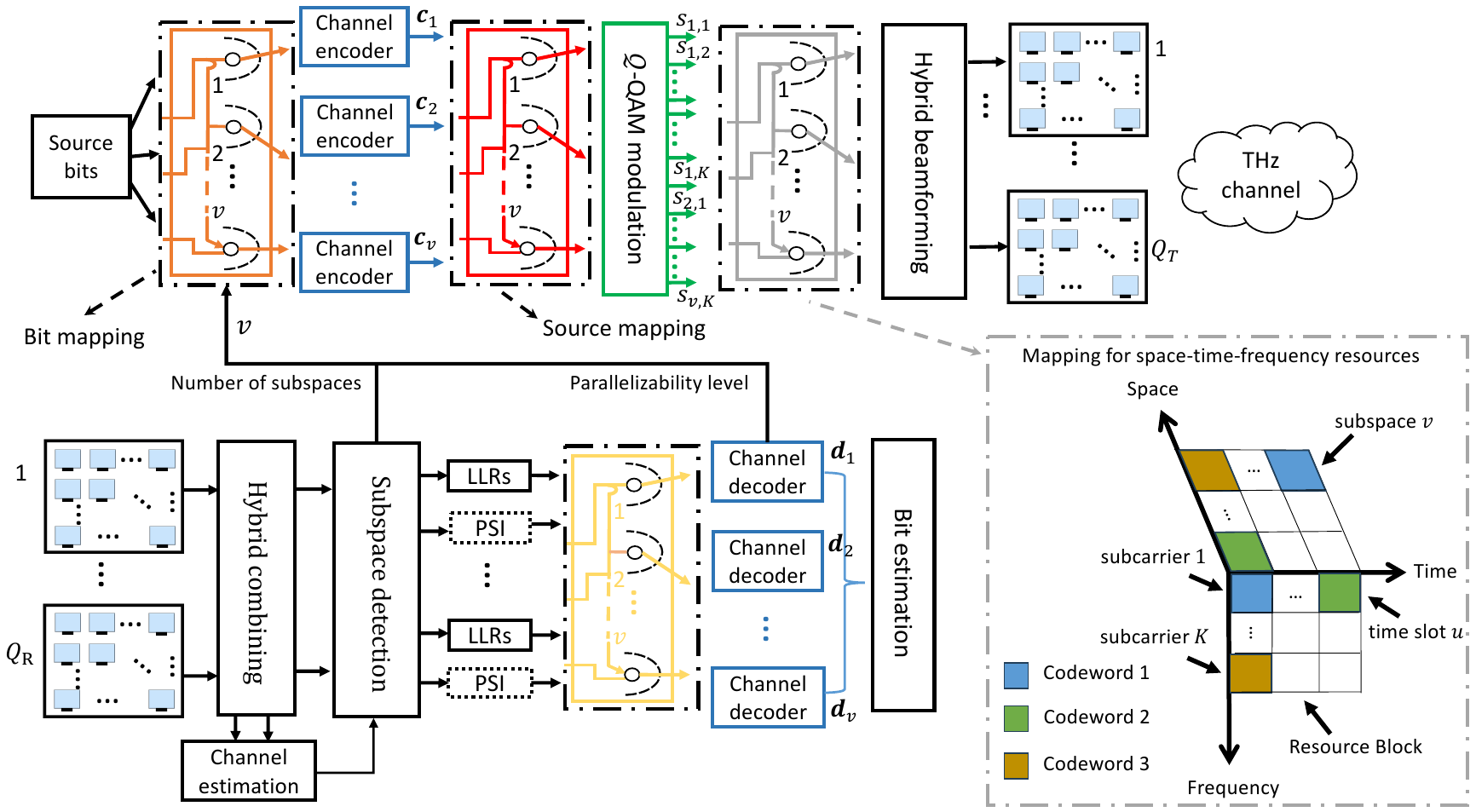}
  \caption{Block diagram of a wideband UM-MIMO THz communication system adopting an AoSA architecture with source parallelizability and pseudo-soft information.}
  \label{fig:aosa_tx_rx}
\end{figure*}

The proposed framework (highlighted in the block diagram of Fig.~\ref{fig:aosa_tx_rx}) comprises two main components: Introducing parallelizability at multiple levels in the baseband (Sec.~\ref{sec:parallel}) and generating pseudo-soft information (PSI) to reduce computational complexity (Sec.~\ref{sec:psi}). The first component involves mapping bits to communication resources (in time, frequency, and space) to enable parallel processing across the entire baseband chain--starting from the source. We leverage the THz channel structures in the second component to generate marginal quasi-static PSI, enhancing the channel-code decoding efficiency~\cite{sarieddeen2022grand}. The study considers coding schemes like turbo, low-density parity-check (LDPC), and polar codes~\cite{balatsoukas2016hardware,ryan2009channel} as candidates for future generations of wireless communications, evaluating their decoders based on key performance indicators including error rate, latency, complexity, and power consumption. The paper also advocates for noise- and interference-centric solutions and explores other research directions (Sec.~\ref{sec:future}).

\section{Prospects and challenges of THz-band, Tbps communications}
\label{sec:tbps_THz}

We first explore THz communications' prospects and challenges, emphasizing the unique THz channel structures that we leverage in our proposed framework. We then introduce the Tbps constraints, stressing the necessity for a revolutionary baseband signal processing approach.

\subsection{THz channel characteristics}
Recent channel measurement campaigns advanced THz channel modeling~\cite{sheikh2022thz}. THz signals undergo significant path loss, comprising spreading and molecular absorption losses~\cite{sarieddeen2021overview}. Higher gas mixing ratios and densities lead to stronger and wider molecular absorption effects that increase with distance, resulting in distance-dependent frequency selectivity, even in line-of-sight (LoS) scenarios. Nevertheless, the available absorption-free transmission spectral windows still cover vast bandwidths. THz channels also demonstrate extreme sparsity in both the time and angle domains, surpassing that of mmWave frequencies. Communication via non-LoS paths is feasible based on current measurement results \cite{sheikh2022thz}. In indoor THz environments, the presence of non-LoS paths is typically limited, further reducing with the integration of massive arrays or high-gain antennas \cite{sheikh2022thz}. However, achieving accurate channel modeling for all scenarios necessitates additional measurements and verification, particularly for frequencies above $\unit[0.45]{THz}$.

The distinctive THz channel characteristics, large bandwidths, and massive antenna numbers present both opportunities and constraints for system design and baseband signal processing. In favorable far-field near-static UM-MIMO THz environments, channels show high correlation in space, time, and frequency, where the small delay spread under high antenna directivity increases the coherence bandwidth and the likelihood of flat fading. However, ideal UM-MIMO signal processing assumes uncorrelated channels, emphasizing the need for THz-specific solutions. Hybrid beamforming emerges as a promising solution in various THz transceiver architectures like sub-connected array-of-subarrays (AoSA) and dynamic subarray (SA) with fixed true-time delay. The AoSA architectures divide UM-MIMO arrays into SAs, each fed with an exclusive radio frequency chain, reducing power consumption and hardware complexity, and simplifying channel estimation compared to a fully digital architecture. Spectral efficiency may decrease compared to fully-connected structures, but with ample bandwidths, enabling multi-user and multiple-stream communications while respecting hardware constraints takes precedence~\cite{sarieddeen2021overview}. 

%Such splitting reduces the computation requirements and results in the processing of less data which alleviates the constraints on baseband processing. 

\subsection{Tbps constraints}

Novel and efficient signal processing is vital to realize Tbps data rates in THz systems.
Achieving a Tbps, even with cutting-edge circuit technologies, requires high parallelism. For instance, considering a very large-scale integration (VLSI) clock frequency of 1\,GHz, 1000 information bits must be processed in parallel when processing 1\,Tbps data. The EPIC project (enabling practical wireless Tbps communications with next-generation channel coding)~\cite{epic2019} aimed at optimizing channel-code decoders by investigating diverse parallelizability techniques and benchmarking results to select relevant key performance indicators. Some resolutions under realistic implementation constraints were to achieve an area efficiency of \unit[100]{gigabit-per-second (Gbps)/$\mathrm{mm^2}$}, using relatively large code-word lengths ($N\!=\!1024$, $N\!=\!2048$), at \unit[$\sim$1]{pico-joule (pJ)/bit} energy efficiency, and \unit[0.1]{Watt per square millimeter (W/$\mathrm{mm^2})$} power density, assuming a clock frequency of \unit[1]{GHz}. Nonetheless, such KPIs are limited by current market technologies; even under idealistic process technology scaling to \unit[7]{nm} nodes or below. Taking into account recent advances in semiconductor transmission capabilities can bring new insights, but with the diminishing effect of Moore’s Law, silicon scaling would provide limited improvements in baseband computations and chip power density. We argue that considering THz-specific factors and seeking parallelism across all baseband blocks is the way forward for THz-band, Tbps signal processing.

\begin{table*}[t]
    \centering
    \small
    \caption{Key characteristics of THz candidate coding and decoding schemes}
\begin{tabular}{|p{1.6cm}|p{7cm}|p{7.7cm}|}
\hline & Prospects                                    & Challenges \\ \hline
Turbo/MAP & \begin{tabular}[c]{p{7cm}}- High code-length flexibility \\ - Adaptive stopping criteria in iterative decoding \end{tabular}  & \begin{tabular}[c]{p{7.5cm}}- Code-rate flexibility\\ - Performance with small code rates\\ - Complexity of iterative decoding\end{tabular}  \\ \hline
LDPC/BP & \begin{tabular}[c]{p{7cm}} - High throughput capabilities\\ - Inherently parallelizable architectures\\ - Good performance with high code rates\end{tabular} & \begin{tabular}[c]{p{7.5cm}}- Encoding complexity\\ - Performance with short code lengths\\ - Code-rate flexibility\end{tabular}  \\ \hline
Polar/SC-List& \begin{tabular}[c]{p{7cm}} - Low encoding complexity \\ - High code-rate flexibility \\ - Favored for short-code, high-code-rate scenarios \end{tabular} & \begin{tabular}[c]{p{7.5cm}} - Decoding complexity\\ - Decoding memory and power consumption\\ - Performance with short and medium code lengths\end{tabular} \\ \hline
Any/GRAND  & \begin{tabular}[c]{p{7cm}} - Universal decoding of arbitrary block codes \\ - Readily parallelizable hardware implementations \\ - Superior in short-code, high-code-rate scenarios \\ - Factors THz-specific channel and noise statistics \end{tabular} & \begin{tabular}[c]{p{7.5cm}} - Complexity in low-code-rate large-code-length scenarios \\ - Random runtime in guessing without abandonment \end{tabular} \\ \hline
\end{tabular}
\label{Table-coding}
\end{table*}

Channel coding and decoding schemes significantly impact system complexity, energy consumption, and overall performance. Turbo, LDPC, and polar codes are contenders for future communication system standards (Table \ref{Table-coding} summarizes their prospects and challenges). Turbo coding is data-flow-based, incorporating limited locality with randomness through interleaving. However, high throughput, high power/energy efficiency, and low latency require highly parallelizable architectures that guarantee significant locality with minimum data flow control. 
%Furthermore, common turbo code designs often adhere to a \cs{I do not think that this is accurate; puncturing enables one to achieve nearly arbitrary granularity in terms of data rates; I suggest removing this} default code rate of 1/3, constraining design flexibility. 
The degree of parallelizability in turbo decoders is thus limited and is highly dependent on the interleaver type, limiting the achievable throughputs compared to LDPC or polar decoders. Although common turbo code designs often adhere to a default rate of 1/3, puncturing mechanisms enable achieving nearly arbitrary granularity in terms of data rates, but at non-negligible performance and complexity costs.
%\green{yes indeed, the interleaver is a limiting factor and for parallelizability, unrolling and pipelining architectures are required in most cases. As for the throughput, the iterative decoding and the hardware implementation complexity compared to LDPC for example creates this gap between Turbo and LDPC in terms of throughput. There is also the code design…}
Polar coding, on the other hand, is of relatively low complexity and is particularly favored in short-code, high-code-rate scenarios. However, polar codes' successive cancellation (SC) decoding is inherently serial, which limits its applicability for high-throughput applications and results in high power consumption. Polar codes generally exhibit good flexibility in code rates. Furthermore, LDPC codes perform well with high code rates but not when applied to short codes, necessitating complex solutions over concatenated architectures. Such solutions come at the cost of heightened complexity and increased energy consumption. The inherent parallelizability of belief propagation (BP) in LDPC decoding favors low-complexity hardware implementations and low power consumption. However, the corresponding code rate flexibility is somewhat limited, despite some clever puncturing schemes demonstrating different rates in fifth-generation (5G) use cases. 
%\cs{5G uses some clever puncturing schemes for providing many different rates; I think that this statement is OK but might not be necessary}
%\green{The cost of puncturing is in the bandwidth efficiency and the additional cost requirements for the specific puncturing techniques. If these can be proven mitigated under THz, then yes we can neglect the code rate flexibility. However, it is still not a certainty.}

%Polar codes generally exhibit good flexibility in code rates, but accommodating varying code rates in successive cancellation decoding would require complex memory management to handle multiple lists of operations. 
%\yellow{Discussion on GRAND is remaining and one additional advantage of Turbo codes related to THz}

%the output nature, and the number of iterations required to achieve a certain level of BER, it can be argued that a tradeoff or an intelligent adjustment of the system parameters is a promising direction. 

Latency reduction is also crucial alongside parallelizability to enable Tbps data rates. Often overlooked in literature is the influence of latency on wireless receiver storage capacity. Let $\Theta$ denote throughput in information bits per second and $L$ represent the latency (in seconds) from analog-to-digital conversion to the output of the physical layer processing engine. The minimum number of bits, $B$, that the physical layer must store is bounded by $\Theta L \leq B$. Storage demands already occupy a significant portion of the baseband processing area for sub-6-GHz and mmWave systems. This issue is projected to become a major bottleneck for Tbps THz systems. For instance, achieving one Tbps throughput with one millisecond processing latency necessitates at least one gigabit storage within the baseband accelerator engine. Apple's M2 Ultra has a second-level cache of 768 megabytes, satisfying this substantial requirement. However, this simple lower bound significantly underestimates actual storage needs as it ignores redundancy (e.g., caused by pipelining, temporary buffers, or redundant data representation). To illustrate this issue, soft decoding uses log-likelihood ratio (LLR) values with five- or seven-bit precision to represent a coded bit. Assuming a rate 1/2 code, this alone renders the bound around ten times lower than actual storage needs. Moreover, multiple parallel data streams are typically acquired at the analog-to-digital converters (ADCs), potentially introducing even more redundancy than the LLR example. This work's proposed framework aims to reduce latency by combining parallelization with short codes, while also cutting down storage by replacing LLRs with PSI.

Another concern arises from the input-output bottleneck in integrated circuits. Consider generating one Tbps of data; the challenge is to efficiently transfer this data off the chip, especially when the chip handles PHY and potentially MAC processing. Recent advances in circuit design have facilitated the development of a 224-Gb/s SERDES receiver in a $\unit[5]{nm}$ FinFET process \cite{9930342}. Even with such throughput advances, the utilization of five of these links would be necessary for handling Tbps data, resulting in significant area overhead, increased power consumption (primarily attributed to the SERDES transceivers), and presenting challenges for packaging, as well as printed circuit board (PCB) or interposer design.

\section{Baseband parallelizability}
\label{sec:parallel}

From the source onwards, hardware and algorithmic efficiency are crucial throughout the communication chain. We argue that using short codes is a prerequisite for achieving such efficiency in Tbps systems, promoting parallelizability, lower latency, and higher throughput.  We propose achieving ``baseband'' parallelizability by intelligently mapping source bits into streams that match target signal processing parallelizability levels. We also highlight the significance of jointly parallelizing MIMO data detection and channel-code decoding processes.

%low-complexity detection schemes for THz is analyzed and the performance of subspace detection schemes that are advantageous in terms of parallelism and hardware efficiency are looked into . Through subspace detection, certain computations can be done in parallel, and thus decoding can occur without significant latency issues. If the mapping of bits and antennas is done accordingly, one can target short and specific code lengths and thus the whole system can see its complexity and global latency reduced.

%Towards bridging the Tbps gap in baseband signal processing, Low latency and complexity, high energy efficiency, and throughput are must requirements of data detection and decoding algorithms and architectures. 

\subsection{Source parallelizability}

We propose a method that introduces parallelizability at the source with structured bit mapping--not random bit interleaving--in space, time, and frequency dimensions. Source parallelizability requires a dynamic baseband chain that enables adaptability in parallelizability attributes across components. In one component, mapping source bits to spatial MIMO architectures is an antenna selection problem that improves the reliability of wireless communication links. The block diagram in Fig. \ref{fig:aosa_tx_rx} illustrates the various elements of source parallelizability, where spatial parallelism is combined with subspace MIMO detection to achieve multiple independent parallel streams of computation. The three illustrated code-words are each composed of two resource blocks represented in a distinct color. We seek efficient mapping of code-word blocks to the available space, time, and frequency resources in a method that enhances the quality of PSI for decoding.

The block diagram of Fig.~\ref{fig:aosa_tx_rx} also illustrates how source parallelizability is complemented with shorter code-words, independently coded over the decoupled streams. Short codes can indeed reduce complexity and latency (consequently, storage requirements) and enhance power efficiency. For example, the practical advantages of short codes are featured in results on ASIC implementations of successive cancellation decoders for polar codes in \cite{dizdar2016high}. The decoder complexities are shown to scale linearly with code lengths; decoding a code-word of length $N$ is $v$-times more complex than decoding a code-word of length $N/v$. This observation is supported by cell count and area measures in decoder designs. The work in \cite{kestel2020506gbit} further illustrates that for polar coding with $N=1024$, $K=518$, throughput is \unit[506]{Gbps}, latency is \unit[309.7]{ns}, and total power consumption is \unit[4.74]{W}. With smaller $N=128$, $K=70$, throughput falls to \unit[64]{Gbps} (without parallelizability), latency to \unit[41.8]{ns}, and power consumption to \unit[0.25]{W}. This example reveals latency reduction and power efficiency gains also in the order of the parallelizability level, $v=10$. 

%The latter can happen by an intelligent allocation of antennas, bit streams, and parallel data detection streams. The performance of such a design heavily depends on the mapping technique.

%Simulation to be added: Rates 0.5 0.75 0.83 0.9, and code lengths 64 128 256 512, polar with SCL L = 8

\subsection{Subspace detection}

For spatial parallelism, we adopt variations of channel subspace decompositions \cite{sarieddeen2017large}. By intelligently encoding and mapping bits to specific antennas, subspace detection schemes allow independent information detection and decoding over spatial streams, effectively reducing global complexity and latency \cite{JemaaDetection2022}. We further note that a subspace detection mechanism is inherently parallelizable. Channel decomposition is often realized through channel-matrix puncturing that breaks dependencies in spatial streams, reducing computations in subsequent conventional detectors. Following subspace matrix puncturing, the data detection complexity can be reduced to approach that of linear zero-forcing (ZF) detectors. 

Channel puncturing typically incurs a diversity loss, especially with rich channels at lower frequencies. Such losses could render forward error correction only helpful in combating noise but not dealing with fading. Although beamforming can counteract such losses, in multi-user scenarios, one would need scheduling (or power-multiplexing) mechanisms to ensure that if two users are on similar beams (locations), then these users should not be served simultaneously. Conversely, performance enhancement is observed under correlated THz MIMO channels with subspace detection (SSD) \cite{JemaaDetection2022}; near-maximum-likelihood detection performance is noted, and complexity reduction comes as a free lunch. THz channels are inherently low-rank in point-to-point MIMO links. Hence, subspace decompositions will not reduce diversity further; they reduce error propagation in back substitution operations of ZF with decision feedback, for example, and result in performance gains.

It is important to note that coupling subspace decomposition with smaller code lengths, although beneficial in terms of complexity, may require proper mitigation to prevent performance degradation. Reducing the code-word size will further aggravate diversity loss, where code size reduction fundamentally causes performance loss, especially when error correction is not across a channel use but per sub-channel. However, the purpose is not to reduce code lengths arbitrarily but to post a tradeoff between performance, overall complexity, and other considerations such as latency and power consumption. The motivation behind short code-words is thus mainly complexity and latency reduction. Extracting PSI from the channel can mitigate such losses, as will be illustrated in Sec. \ref{sec:psi}.

%On one side, we have the complexity reduction compared to ML, and from the other the high parallelizability structure, with near ML performance in certain channel conditions. 

\begin{figure}[t]
  \centering
  \includegraphics[width=0.48\textwidth]{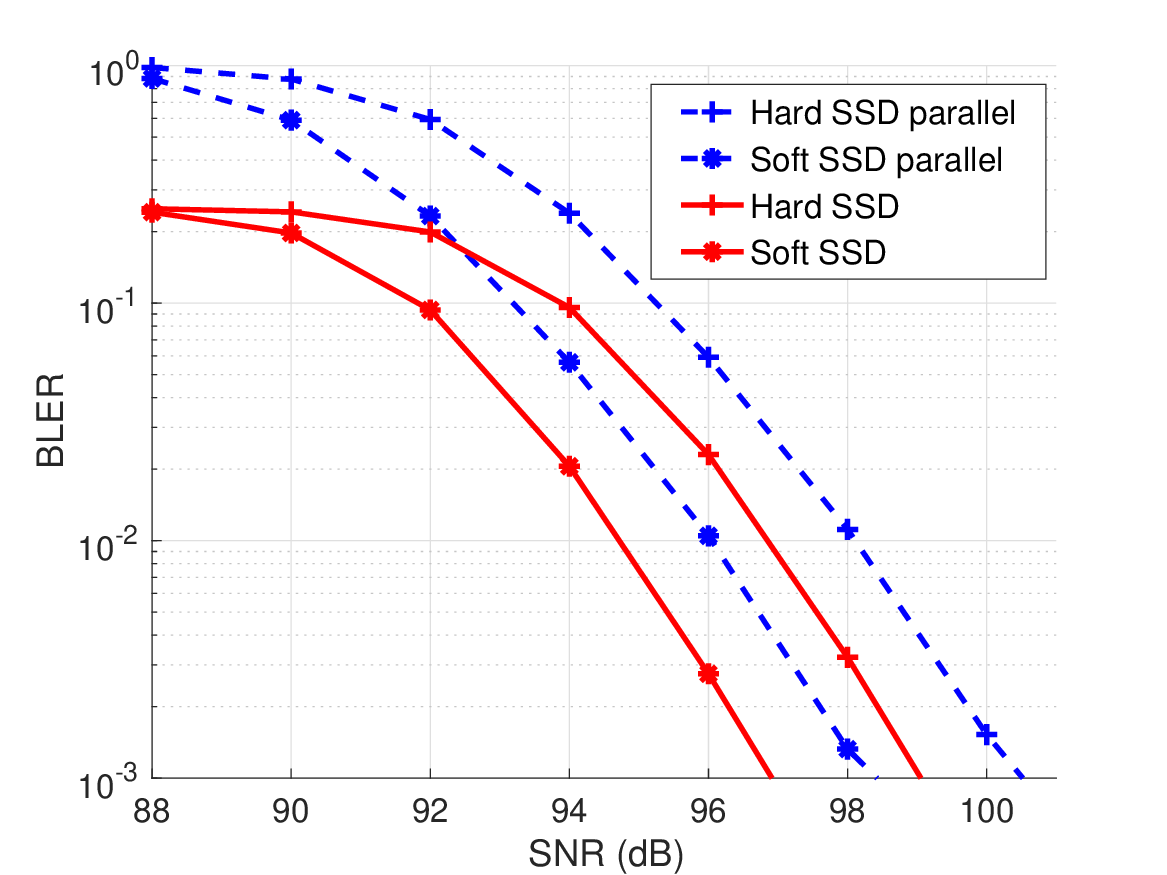}
 \caption{BLER versus SNR of hard- and soft-decoded systems under SSD and polar coding, with parallelizability ($N=128$, $K=74$) and without parallelizability ($N=512$, $K=296$) - QPSK - 4x4 MIMO, SC-List decoding ($L=16$) - frequency selective THz channels.}
  \label{fig:puncturing}
\end{figure}
We assess the effectiveness of the proposed parallelizability framework in the context of polar codes and contrast it with a traditional non-parallel architecture. Figure \ref{fig:puncturing} illustrates block error rate (BLER) results of hard and soft decoding with SSD for a frequency-selective $\unit[0.3]{THz}$ channel generated by our TeraMIMO simulator (which we also adopted in~\cite{JemaaDetection2022}). The simulation chain exhibits $v$ independent and parallel streams. Each of these streams is CA-polar encoded with a code rate close to 2/3 ($N\!=\!128$, $K\!=\!74$). A benchmark non-parallel architecture with $N\!=\!512$ and $K\!=\!296$ is also simulated. The results demonstrate a performance discrepancy under parallelizability of around $\unit[1]{dB}$ in soft and hard decoding at BLER of $10^{-3}$. Such a performance loss is graceful, considering the benefits of shorter code lengths. The gain in soft decoding compared to hard decoding, however, is limited to $\unit[2]{dB}$ because we average over a relatively large dynamic range of practical frequency-selective THz channels (over a $\unit[5]{GHz}$ bandwidth).

% figure on the distributions of the R matrix and punctured R matrix and compare them to the Rayleigh channel

\subsection{Decoder parallelizability}

In addition to parallelizing the baseband from the source, channel-code decoding should possess sufficient spatial (in terms of area) and functional parallelism and large data locality and structural regularity. Several techniques have been explored in the literature to enhance the parallelizability of THz candidate coding/decoding schemes. One such technique is unrolling, applied to turbo decoding in architectures like cross-maximum a posteriori (MAP) \cite{epic2019} to increase throughput. However, structuring the decoder iterations in a chain architecture fixes the number of iterations bounding the pipeline stages, which can limit the performance gains and power-saving abilities. On the other hand, LDPC codes are inherently parallel; LDPC decoders utilize the BP algorithm, in which check and variable node computations can be executed independently. The throughput capabilities of a BP-based LDPC decoder depend on how many edges are processed in parallel. Moreover, pipelining and unrolling are used together to improve the level of parallelizability in BP, breaking data dependencies and further improving performance. As for Polar codes, although SC-List has low implementation complexity, the decoding process itself is intrinsically sequential, imposing constraints on the maximum achievable throughput. An avenue to enhance throughput lies in the potential unrolling of the tree traversal. This approach yields a data-flow architecture amenable to pipelining and unrolling.

Under the $\unit[28]{nm}$ technology, remarkable advances have been achieved in the implementation of main coding schemes, paving the way for prospective THz/Tbps applications~\cite{epic2019}. LDPC codes, using unrolling and pipelining with a block size of 672 and a code rate of 13/16, attained an energy efficiency of $\unit[6]{pJ/bit}$ and throughput of $\unit[160]{Gbit/s}$. Similarly, polar codes, operating with a block size of 1024 and a code rate of 1/2, exhibited a throughput of $\unit[620]{Gbit/s}$ at a power consumption of $\unit[2.8]{W}$ and an energy efficiency of $\unit[4.6]{pJ/bit}$. As for turbo codes, a block size of 192 permitted a throughput of only $\unit[48]{Gbit/s}$ and an area efficiency of $\unit[5.33]{W/mm^2}$.

\section{Leveraging pseudo-soft information in THz system design}
\label{sec:psi}

We argued that source parallelism mechanisms and the use of short code lengths, although they reduce complexity and latency, can result in performance degradation. We propose to mitigate such degradation with low complexity overhead by leveraging THz channel characteristics in soft decoding information. However, generating such soft information in the form of LLRs can be prohibitively complex under Tbps constraints. Furthermore, passing high-resolution LLRs to the channel decoder in each channel use can create a bandwidth bottleneck in interconnects between baseband modules; LLR storage is also challenging. Leveraging the LLRs in iterative soft detection and decoding techniques, while potentially advantageous in mitigating THz channel effects and enhancing BLER performance, is also discouraged under THz-band, Tbps constraints. Iterative schemes multiply complexity and memory issues, leading to adverse effects on throughput and latency.

To improve performance while addressing these concerns, we propose incorporating channel state information (CSI) and additive noise statistics into channel bit mapping and code design, then leveraging CSI in PSI for a single-shot (no iterations) data detection and decoding mechanism. PSI can be expressed as the effective signal-to-noise ratio (SNR) following detection processing \cite{sarieddeen2022grand}, and it only depends on the channel. Such information can thus be computed once over a channel coherence time/bandwidth, significantly reducing complexity. Per realization, only hard-output bits will be passed from the detector to the decoder. The need for generating per-bit soft-detection reliability information in both linear and non-linear detectors is elevated at a graceful performance cost when PSI is rich. PSI richness is related to bit mapping design, where combining bits from different transmission sources in a single code-word is favored. Additionally, leveraging the channel structure in PSI eliminates the need for costly bit-interleaving and noise-whitening operations.

%The concept of PSI was initially introduced in~ for linear detectors only, but this work extends it to non-linear detectors as well.

We thoroughly assess the suitability of the resulting PSI for the THz system design with different channel decoding schemes. Our investigation includes a range of low-complexity data detectors for PSI generation. Among the linear equalizers, we explore zero-forcing (ZF) and minimum mean square error (MMSE) techniques, and for the non-linear equalizers, we examine the chase detector (CD) based on QRD decomposition. We also explore its punctured version, namely punctured chase detector (PCD), which utilizes WRD decomposition \cite{sarieddeen2017large}.
To generate the PSI for these non-linear detectors, we propose using the $\mbf{R}$ matrix resulting from decomposing the THz channel using QRD (and punctured $\mbf{R}$ when using WRD), along with the noise variance. For each layer, the PSI is obtained by summing the squares of corresponding $\mbf{R}$ rows' elements.

\begin{figure}[t]
  \centering
  \includegraphics[width=0.48\textwidth]{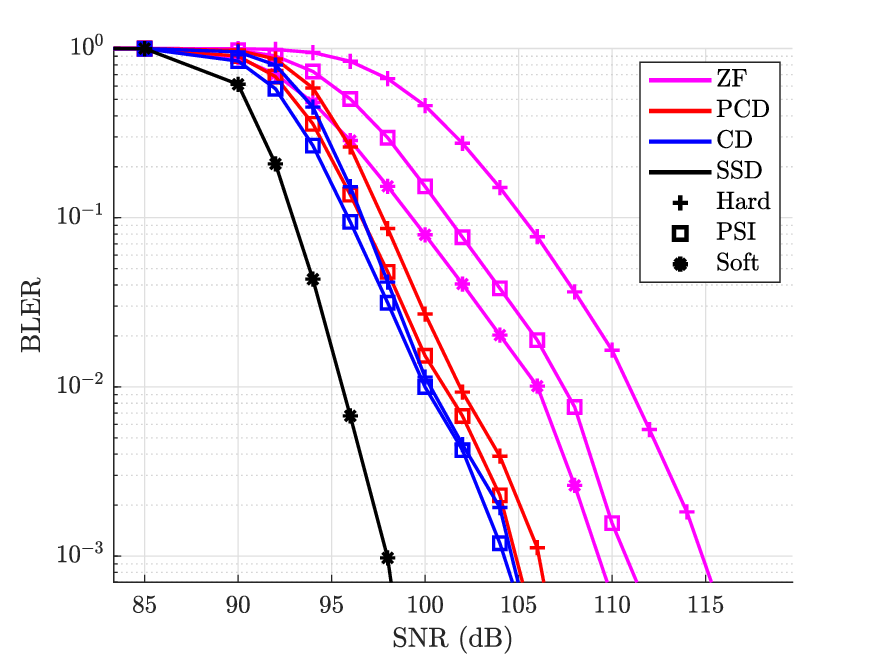}
 \caption{BLER versus SNR for hard-, PSI-, and soft-decoded parallelizable systems for different linear and non-linear detection schemes - QPSK modulation - turbo coding ($K\!=\!256$ and $R\!=\!1/3$) - frequency selective THz channels.}
  \label{fig:turbo_thz}
\end{figure}

\begin{table*}[htb]
    \centering
    \small
    \caption{Complexity and latency measures in the proposed framework}
\begin{tabular}{|c|c|c|c|}
\hline
   & Complexity & Performance   & Latency     \\ \hline
CD hard & $\mathcal{O}(|\mathcal{X}|M^3) +C_{\mathrm{Dec}}$          & Low reliability  & $L_{\mathrm{Det}}+L_{\mathrm{Dec}}$  \\
\hline
PCD hard & $\mathcal{O}(|\mathcal{X}|M^3-\theta_1) + C_{\mathrm{Dec}}$ & Low reliability & $(L_{\mathrm{Det}}-\eta_{1})+L_{\mathrm{Dec}}$ \\
\hline
PSI-aided PCD & $\mathcal{O}(|\mathcal{X}|M^3-\theta_1)+ \theta_2C_{\mathrm{Dec}}$  & Approaches soft  & $(L_{\mathrm{Det}}-\eta_1)+\eta_2 L_{\mathrm{Dec}}$ \\
\hline
PCD soft & $\mathcal{O}(\alpha_{\mathrm{LLR}}(|\mathcal{X}|M^3-\theta_1))+ \beta\theta_2C_{\mathrm{Dec}}$ & Best PCD performance & $\Bar{\alpha}_{\mathrm{LLR}}(L_{\mathrm{Det}}-\eta_1)+\beta \eta_2 L_{\mathrm{Dec}}$ \\
\hline
\begin{tabular}[c]{@{}c@{}}PSI-aided PCD with \\ source parallelizability\end{tabular} & $\mathcal{O}(|\mathcal{X}|M^3-\theta_1)+\theta_2 C_{\mathrm{Dec}}/v$ & \begin{tabular}[c]{@{}c@{}}Slight degradation \\ compared to PSI \end{tabular} & ${(L_{\mathrm{Det}}-\eta_1)}+\eta_2 L_{\mathrm{Dec}}/v$ \\ \hline
\end{tabular}
\label{table:complexity}
\end{table*}

To evaluate the performance of THz PSI and the proposed parallelizability framework, we conduct comprehensive simulations and highlight sample results in Fig.~\ref{fig:turbo_thz}. We compare the BLER versus the SNR for hard-, PSI-, and soft-decoded systems. In this simulation setup, we also adopt an indoor scenario with a carrier frequency of $\unit[0.3]{THz}$ and a bandwidth spanning $\unit[5]{GHz}$. Further details, encompassing multi-path components and other THz-related parameters, are configured in the TeraMIMO simulator. The system consists of $4\!\times\! 4$ SAs, utilizing a single-carrier waveform over multiple separated frequency bands, a scheme recognized for its advantages in THz communications~\cite{sarieddeen2021overview}. In this setup, each symbol experiences a different channel gain, i.e., rich diversity, due to frequency-selectivity.

The results in Fig.~\ref{fig:turbo_thz} illustrate that both CD and PCD outperform ZF. Adopting PCD instead of CD with THz channels results in significant complexity reduction and allows parallelism; the corresponding loss in BLER is graceful ($\unit[1]{dB}$). PSI-decoded systems outperform hard-decoded systems by multiple dBs, where in the ZF case, the gain is $\unit[4]{dB}$, measured at BLER$\ =10^{-3}$, and it reduces to $\unit[1.2]{dB}$ and $\unit[0.4]{dB}$ for PCD and CD, respectively. We add to Fig.~\ref{fig:turbo_thz} a reference soft SSD detector, which forms a lower performance bound for both CD and PCD. The corresponding significant $\unit[6.5]{dB}$ performance gap comes at the cost of computing soft-output LLRs following multiple parallelizable subspace decompositions. Nevertheless, with ZF detection, PSI-assisted decoding approaches soft decoding, with a performance gap of about $\unit[1.2]{dB}$. These results further assure the validity of adopting PSI in the design of low-complexity non-iterative THz detectors. It is noteworthy to emphasize that when integrating PSI with correlated channels (even in time-domain), such as in far-field LoS flat-fading scenarios, the performance mirrors that of hard decoding due to the absence of channel/PSI diversity.

The results in Fig.~\ref{fig:grand_thz} further illustrate the performance of the proposed parallelizability framework under the same THz indoor channel conditions with SC-List and GRAND decoding of polar codes ($K/N\!=\!116/128$), assuming ZF detection in a $4\!\times\!4$ MIMO system. PSI captured most of the structure in the channel, resulting in a marginal gap of $\unit[2]{dB}$ compared to soft decoding. GRAND further outperforms SC-List decoding by $\unit[2]{dB}$ as it jointly considers CRC and polar code redundancies. Interestingly, the complex soft SC-List decoding (with a list size of $16$) is matched by low-complexity PSI-based GRAND decoding, further highlighting the importance of using noise-centric decoders in this range of high rates and short codes.

%Moreover, the performance comparison illustrated in Fig.~\ref{fig:polar_thz} shows the validation of the PSI for MMSE detector with short-codes high-rate polar decoding scheme under the proposed parallelizability framework. The performance of the PSI version outperforms the hard version with about $\unit[1.4]{dB}$, measured at BLER$\ =10^{-3}$. However, in the case of the PCD detector, the performance is almost identical to the hard version. The previous result opens the question about optimal PSI design for different decoding schemes such as Polar with high rates.

%\begin{figure}[t]
%  \centering
%  \includegraphics[width=0.48\textwidth]{Magazine_PCD_MMSE_SSSD_POLAR_THz.eps}
% \caption{Comparison of the BLER versus SNR of hard, PSI, and soft coded systems for different linear and non-linear detection schemes, QPSK modulation, and polar code (with number of uncoded bits is $105$ and a rate $R=0.82$) by using frequency selective THz channel using TeraMIMO simulator and adopting the proposed parallelizability framework.}
%  \label{fig:polar_thz}
%\end{figure}
\green{
\begin{figure}[t]
  \centering
  \includegraphics[width=0.48\textwidth]{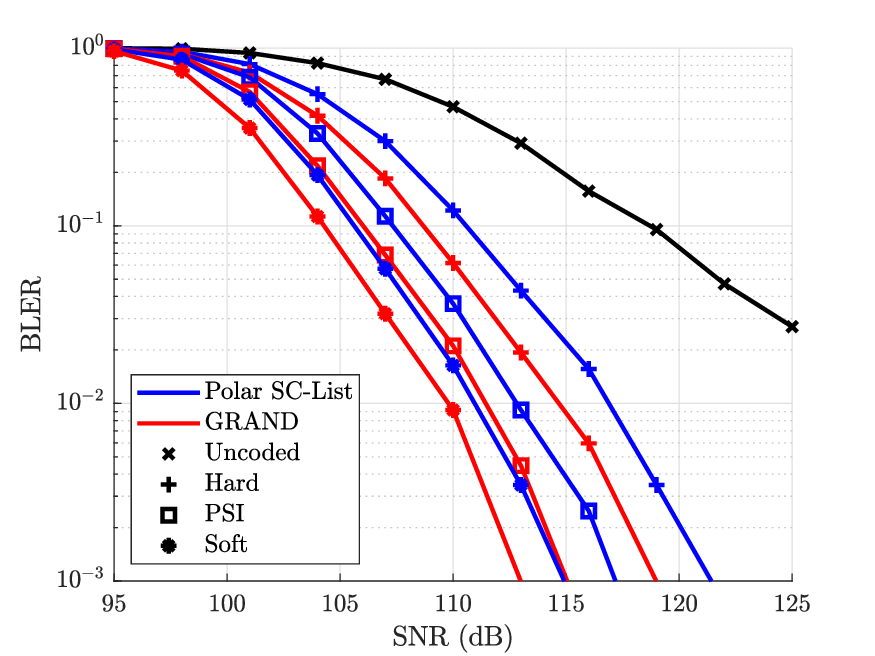}
 \caption{BLER versus SNR for uncoded-, hard-, PSI-, and soft-decoded parallelizable systems with ZF detection - QPSK modulation - SC-List ($L=16$) and GRAND decoding of polar codes ($N=128$, $K=116$) - frequency-selective THz channels.}
  \label{fig:grand_thz}
\end{figure}
}
Table \ref{table:complexity} highlights the complexity, performance, and latency of variations of the studied chase detectors in terms of the $\mathcal{O}$ notation that describes the upper bound of an algorithm's complexity in terms of input size. We consider, without loss of generality, a symmetric UM-MIMO system where the number of transmit and receive antenna SAs is $M$. The detection complexity is expressed in terms of $M$ and the cardinality of the modulation constellation, $|\mathcal{X}|$. The CD mitigates error propagation in SIC by an exhaustive search over the root layer. The PCD executes the search over a transformed punctured channel; the puncturing WRD incurs additional processing. Following puncturing, fewer detection operations are performed over a more sparsely decomposed channel. This reduction in the number of operations reduces the CD complexity, $\mathcal{O}(|\mathcal{X}|M^3)$, by a factor of $\theta_1$ (can reach up to $50\%$ \cite{sarieddeen2017large}) while diminishing the detector latency, $L_{\mathrm{Det}}$, by a factor of $\eta_1$. 

The more significant reduction in complexity and latency is achieved via source parallelizability across the baseband chain and using short codes. By considering $v$ to be the degree of achieved parallelizability (the number of independent computation streams effectively parallelized through proper source bit mapping), the decoding complexity, $C_{\mathrm{Dec}}$, and latency, $L_{\mathrm{Dec}}$, can be divided by $v$ (extrapolations can be made for power consumption). Furthermore, under PSI-aided PCD, the decoding complexity and latency also decrease by multiplicative factors $\theta_2\!<\!1$ and $\eta_2\!<\!1$, respectively, because of enhanced computational efficiency compared to hard decoding. Even higher reductions are expected with soft decoding with optimal LLRs ($\beta\theta_2$ and $\beta\eta_2$; $\beta\!>\!1$), but at an additional cost of LLR computations in detection, $\alpha_{\mathrm{LLR}}\!>\!1$. PSI generation, however, requires negligible additional computations compared to hard detection. 

\section{Research challenges and opportunities}
\label{sec:future}

This last section highlights several future research directions for leveraging information on THz channel and noise in low-complexity baseband processing.

\subsection{PSI for hybrid cross-field communication}

Due to the extremely short wavelengths, massive arrays, and limited communication distances, a substantial portion of practical THz scenarios falls within the near-field. Consequently, THz systems are expected to operate within both near-field and far-field regions, leading to the emergence of a novel paradigm termed cross-field THz communications~\cite{tarboush2023cross}. Moreover, the coexistence of far and near-field channel paths is also highly probable, giving rise to the concept of hybrid-field THz communications. The spherical wave model is appropriate for channel modeling, estimation, and baseband signal processing in such cases. Accounting for the near-field THz channel structures provides more spatial degrees of freedom in comparison to far-field channels. Such spatial richness offers novel opportunities for resource bit-mapping and introduces novel dimensions--such as distance--contributing to enhanced source parallelizability. Higher channel diversity is expected across the UM-MIMO near-field channels, which, in our context, translates to richer PSI generation at detector outputs. 

\subsection{Guessing Random Additive Noise Decoding (GRAND)}

Guessing random additive noise decoding (GRAND) recovers code-words by guessing rank-ordered putative noise sequences and reversing their effect until valid code-words are obtained. There are several arguments in favor of adopting GRAND in THz-band, Tbps settings. First, GRAND is a universal decoding mechanism that can decode any block-code construction, enabling low-cost reconfigurable architectures that can support the requirements of diverse emerging THz communications applications. GRAND particularly performs well with short moderate-redundancy codes that naturally arise from parallelizable baseband architectures. GRAND has also demonstrated good performance leveraging PSI on additive-noise statistics and channel-state information in fading channels. The performance of the hardware-friendly ordered reliability bits GRAND (ORBGRAND) under PSI extracted from linear ZF and MMSE equalization approximates the performance of state-of-the-art decoders of CRC-assisted polar (CA-polar) and Bose–Chaudhuri–Hocquenghem (BCH) codes that avail of complete soft information~\cite{sarieddeen2022grand}. 

The more we know about the structure and correlation in THz channel and noise models, the better GRAND can be optimized for THz scenarios. The prospects of GRAND are benchmarked to other decoding schemes in Table \ref{Table-coding}. It is worth mentioning, however, that maximum-likelihood GRAND can result in random runtime as the number of guesses can vary between realizations of channel use. To bound latency, we have to sometimes terminate guesswork early, which is referred to as GRAND with abandonment. Furthermore, other variations of universal noise-centric decoders also offer performance and complexity trade-offs worth investigating in a THz-band, Tbps context, such as variations of ordered statistic decoding (OSD) \cite{9427228}, which has demonstrated good performance in decoding short codes.

\subsection{Other noise sources and noise recycling}

The molecular absorption noise, which constitutes the majority of the noise in wideband THz channels (especially in nano-communication scenarios), is shown to be colored over frequency~\cite{sarieddeen2021overview}. Such noise follows a Gaussian distribution and has varying total noise power based on the composition of the medium, specifically gases and their isotopologues' mixing ratios, as well as other factors such as pressure, temperature, and humidity. Absorption noise could be leveraged as a source of information, combined with PSI, to enhance THz detection and decoding schemes. Better knowledge of noise can also be factored into noise recycling schemes. Noise recycling is particularly favored in scenarios involving multiple channels subjected to correlated noise, and its corresponding mechanism is compatible with GRAND~\cite{Riaz10041405}. Correlated noise also arises in THz UM-MIMO systems, mainly because of mutual coupling effects when antennas are closely spaced. Noise recycling can serve as an alternative to, or in conjunction with, PSI. In general, PSI favors richness in noise, whereas noise recycling favors strong noise correlation. Furthermore, phase noise in THz devices poses more severe consequences than in microwave or mmWave devices. Effectively mitigating such effects demands advanced and complex techniques. Hence, exploiting low-complexity techniques that leverage the possible structure of such impairments, which can be modeled using either correlated or uncorrelated Gaussian models~\cite{sarieddeen2021overview}, is still an open question.

\subsection{The role of AI}
Artificial intelligence (AI) and machine learning are poised to revolutionize THz communications. In our proposed framework, reinforcement learning (RL) can be used to optimize bit mapping strategies across time, frequency, and space resources by enabling agents to interact with dynamic UM-MIMO THz channels, thereby enhancing specific performance metrics like throughput. In addition, recurrent neural networks (RNNs) have emerged as viable replacements for traditional channel decoders and detectors, particularly for non-linear detectors where finding the optimal solution is analytically challenging. RNNs can thus offer practical approaches to obtaining PSI in subspace detection schemes. Furthermore, generative AI could assist in obtaining PSI by learning the statistical properties of errors and noise, especially by leveraging recently proposed diffusion models that rely on the denoising process.

AI tools can further assist in THz channel estimation. Our PSI design assumes perfect CSI, which is very challenging in UM-MIMO systems. Fixed-point networks present a promising solution for efficient UM-MIMO THz channel estimation, offering adaptive complexity and linear convergence guarantees with minimal iterations. Extending such solutions within our framework is crucial. However, it's worth noting that although many AI-driven solutions for THz transceiver design could be explored, our primary objective is to reduce complexity significantly. For example, under high parallelizability, model-based noise-centric decoders can be of much lower complexity and more energy efficient than AI-based solutions.

\section{Conclusions}

This paper addresses the critical challenges and constraints in realizing efficient THz communication paradigms that support Tpbs data rates. We propose an innovative framework for low-complexity THz-band baseband signal processing that fosters parallelizability and leverages quasi-static THz channel structures. Source parallelizability optimizes bit mapping to spatial, temporal, and frequency resources, whereas PSI leverages channel state information and noise statistics for enhanced low-complexity detection and decoding. 
%Finally, We have highlighted the importance of considering both near-field and far-field effects in the design of PSI and source parallelizability in the general hybrid cross-field communication scenarios, explored the potential of integrating AI and machine learning for optimized system design, and introduced potential extensions for noise recycling as an alternative to PSI in the case of channels with correlated noise. These directions collectively pave the way for unlocking the full potential of THz communication systems.

% Generated by IEEEtran.bst, version: 1.14 (2015/08/26)

\break

\section*{Biographies}
\footnotesize

\noindent\textbf{Hadi Sarieddeen} (S'13-M'18) is an Assistant Professor at AUB, Lebanon.
\hfill\break
\hfill\break
\textbf{Hakim Jemaa} (S'22) is a Ph.D. student at KAUST, Saudi Arabia.
\hfill\break
\hfill\break
\textbf{Simon Tarboush} is an independent researcher from Damascus, Syria.
\hfill\break
\hfill\break
\textbf{Christoph Studer} (SM'14) is an Associate Professor at ETH Zurich, Switzerland.
\hfill\break
\hfill\break
\textbf{Mohamed-Slim Alouini} (S'94-M'98-SM'03-F'09) is a Distinguished Professor at KAUST, Saudi Arabia.
\hfill\break
\hfill\break
\textbf{Tareq Y. Al-Naffouri} (M'10-SM'18) is a Professor at KAUST, Saudi Arabia.
\end{document}